\documentclass[aps,twocolumn]{revtex4}
\usepackage{graphicx}
\begin{document} 
\title{Electronic structure and correlations in pristine and potassium doped Cu-Phthalocyanine molecular crystals}
\author{Gianluca Giovannetti$^{1,2}$, Geert Brocks$^{2}$ and Jeroen van den Brink$^{1,3}$}

\address{
$^1$Institute Lorentz for Theoretical Physics, Leiden University,
          P.O. Box 9506, 2300 RA Leiden, The Netherlands \\
$^2$Faculty of Science and Technology and MESA+ Research Institute, University of Twente,
            P.O. Box 217, 7500 AE Enschede, The Netherlands \\
$^3$Institute for Molecules and Materials, Radboud Universiteit Nijmegen, P.O. Box 9010, 6500 GL  Nijmegen, The Netherlands
}

\begin{abstract} 
We investigate the changes in the electronic structure of copper phthalocyanine (CuPc) crystals that is caused by intercalation with potassium. This is done by means of {\it ab initio} LSDA and LSDA+U calculations of the electronic structure of these molecular crystals. Pristine CuPc is found to be an insulator with local magnetic moments and a Pc-derived valence band with a width of 0.32 eV. In the intercalated compound $\rm K_2CuPc$ the additional electrons that are introduced by potassium are fully transferred to the $e_g$ states of the Pc-ring. A molecular low spin state results, preserving, however, the local magnetic moment on the copper ions. The degeneracy of the $e_g$ levels is split by a crystal field that quenches the orbital degeneracy and gives rise to a band splitting of 110 meV.  Molecular electronic Coulomb interactions enhance this splitting in $\rm K_2CuPc$ to a charge gap of 1.4 eV. The bandwidth of the conduction band is $0.56$ eV, which is surprisingly large for a molecular solid. This is line with the experimentally observation that the system with additional potassium doping, $\rm K_{2.75}{CuPc}$, is a metal as the unusually large bandwidth combined with the substantial carrier concentration acts against localization and polaron formation, while strongly promoting the delocalization of the charge carriers. 
\end{abstract}
\date{\today} 
\maketitle

{\it Introduction}
Molecular crystals such as for instance fullerenes, polyacetylene or $\rm (BEDT-TTF)_2X$ salts which are chemically doped show a dramatic change in their electron transport properties~\cite{Ishiguro98,Forro01}. Starting out as insulators, upon doping they can become conducting and in special cases even superconducting.  Because of the chemical richness of molecular crystals,  building new metals by doping them remains an exciting and ever moving front. Very recently the Delft group showed that transition metal phthalocyanines (MPcs) FePc, CoPc, NiPc, CuPc, initially insulating, can be turned into a metal through potassium doping~\cite{Craciun03}.  These molecules are particularly interesting because they have a magnetic moment, which opens the way to interesting electron correlation effects. Because the molecules can have in addition a possible orbital degeneracy, Jahn-Teller coupling, Hund's rule exchange and large on-site Coulomb interactions, the doped metallic molecular solids constitute a novel class of strongly correlated metals~\cite{Tosatti04}. The question arises how upon doping the interplay develops between metallicity, Jahn-Teller distortions and collective correlation effects such as magnetic ordering. We set the stage for more involved many-body calculations by performing {\it ab initio} electronic structure calculations on pristine and potassium doped copper phthalocyanine $\rm K_2CuPc$, which has the structure of the experimentally observed metallic phase with $x=2.75$.  For this doped compound $\rm K_{2.75}CuPc$ we wish to establish: (i) whether the potassium atoms in  donate their electrons to the CuPc, and if so, how many and in which molecular orbitals they occupy (Cu-related or Pc-related), (ii) which are the properties the valence and conduction bands that are derived from these molecular orbitals, and in particular, (iii) whether the bandwidth such that it explains the observed metallic behavior of our doped molecular crystal. This indeed turns out to be the case: the metallicity of $\rm K_{2.75}CuPc$ is due to its relative large bandwidth.

\begin{figure}
\centerline{\includegraphics[width=0.4\columnwidth]{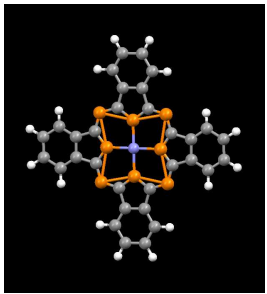}}
\caption{The CuPc molecule with $D_{4h}$ symmetry. It is planar and the central $\rm Cu^{2+}$ ion is surrounded by two rings of each 4 nitrogen atoms. Carbon and hydrogen make up the other atoms in the molecule. 
}
\label{fig:CuPc}
\end{figure}

{\it Pristine CuPC: LSDA} 
The structure of the CuPc molecule and its electronic energy level diagram~\cite{Rosa94,Liao01} is shown in Fig.~\ref{fig:CuPc}. The molecule is flat and has a four-fold rotation symmetry, $D_{4h}$. CuPc is an open shell molecule: the Pc-ring is essentially a closed shell system, but copper is divalent and thus in a $d^9$ configuration, with a single hole that carries an uncompensated magnetic moment. It is important to note that the first electron removal state of the CuPc molecule is a Pc-derived $a_{1u}$ state and the first electron addition state is a Pc-derived $e_g$ state. Also adding a second (and even third) added electron will {\it not} fill the Cu $b_{1g}$ state, but instead be added to the $e_g$ state of the Pc-ring, see Table~\ref{tab:groundstate}. 

\begin{table} 
\label{tabname} 
\centering\begin{tabular}{|l|c|} 
\hline\hline 
Ion  & Groundstate Configuration \\ 
\hline 
 $CuPc^{+}$   & ${(a_{1 u})}^1 {(b_{1g})}^1 {(2e_{g})}^0$    \\ 
 $CuPc$        & ${(a_{1u})}^2 {(b_{1g})}^1 {(2e_{g})}^0$    \\ 
  $CuPc^{-}$   & ${(a_{1 u})}^2 {(b_{1g})}^1 {(2e_{g})}^1$    \\ 
  $CuPc^{2-}$   & ${(a_{1u})}^2 {(b_{1g})}^1 {(2e_{g})}^2$    \\ 
$CuPc^{3-}$   & ${(a_{1u})}^2 {(b_{1g})}^1 {(2e_{g})}^3$    \\ 
\hline\hline 
\end{tabular} 
\label{tab:groundstate}
\caption{Electronic ground state configurations for CuPc with various valencies. The $a_{1u}$ and $e_g$ levels are Pc-derived and the $b_{1g}$ level is a Cu orbital.} 
\end{table} 

We start by investigating the electronic structure of undoped solid CuPc before studying the effects of doping. In its $\beta$-structure the unit cell has a $C_{2h}$ space-group symmetry and there are two CuPc molecules per the unit cell, the first one in the origin of the unit cell while the second molecule at $(a/2,b/2,0)$~\cite{Brown68}. 
The molecules are slightly distorted and their rotation symmetry in the crystal is lowered to a two-fold rotation $D_{2h}$. 

The electronic structure calculations are done within the density functional approach in the local density approximation (LDA) and generalized gradient approximation (GGA) for the intercalated system, using the Vienna Ab-Initio Simulation Package (VASP) \cite{VASP}. We first determined the magnetic groundstate of the compound. From the computations in the SGGA framework we find that the ground state exhibits antiferromagnetic order, and that the energy gain of the antiferromagnetic state with respect to the paramagnetic one is $394$ meV per unit cell. The magnetic moment of each molecule is close to 1 $\mu_B$. However, we also find that the energy difference between the ferro and antiferromagnetic configuration of the two Cu spins in the unit cell is very small, i.e. smaller than 1 meV. This indicates that even though local moments are present in this system, a magnetic ordering temperature --if ordering occurs at all-- is expected to be very low. It also implies that the band structure that we will determine next is nearly identical for the ferro and antiferromagnetically ordered groundstate. In these total energy calculations a plane-basis set with a kinetic energy cutoff of 550 eV was chosen and the tetrahedron method with Blochl correction using 40 irreducibile k-points was used.

\begin{figure}
\includegraphics[width=0.8\columnwidth]{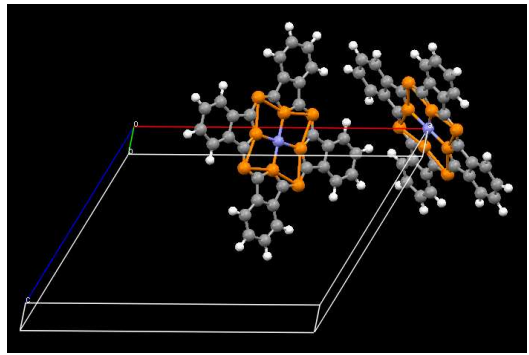}
\caption{Schematic representation of the crystal structure of pristine CuPc. There are two inequivalent molecules in the unit cell.}
\label{fig:structureCuPc}
\end{figure}

\begin{figure}
\includegraphics[width=1.0\columnwidth]{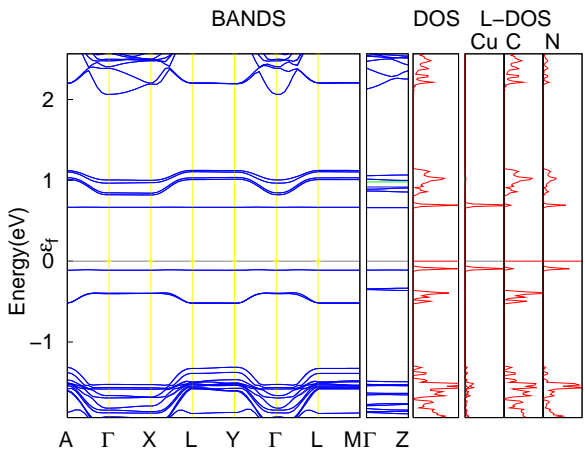}
\caption{
Band-structure, density of states and projected density of states for CuPc, calculated with SGGA.
The energies are plotted along lines in the first Brillouin zone, connecting the points:
The Fermi-level is the zero of energy of the system.
$A = (0.5,0.5,0.0)$,
$\Gamma=(0.0,0.0,0.0)$,
$X =(0.5,0.0,0.0)$,
$L =(0.5,0.5,0.5)$,
$Y=(0.0,0.5,0.0)$,
$M=(0.0,0.5,0.5)$,
$Z=(0.0,0.0,0.5)$.}
\label{fig:CuPc_LSDA}
\end{figure}

The band structure and density of states of CuPc, as calculated by LSDA is shown in Fig.~\ref{fig:CuPc_LSDA}. The bands are arranged in almost degenerate pairs, indicating that the local electronic structure of the two molecules in the unit cell is practically identical. Moreover the bands show little dispersion, from which we conclude that the interaction between the molecules is weak.  In fact there is only appreciable dispersion of the bands along the crystalographic $b$-direction, as can clearly be seen from, for instance, the $\Gamma-Y$  cut through the Brioullin-zone in Fig.~\ref{fig:CuPc_LSDA}. Thus from an electronic point of view, solid CuPc is quasi one-dimensional.

In Fig.~\ref{fig:CuPc_LSDA} we see that below the Fermi level there is a very flat band, which is derived from localized Cu $b_{1g}$ states. Above the Fermi-level is the corresponding empty Cu state. At still higher energy are the empty $e_g$ states. It is interesting to see that in the band structure the $e_g$ states, which are doubly degenerate in the molecule, split up into two non-degenerate bands. The degeneracy is lifted in the crystal because of the ionic and covalent molecular crystal fields that distort the CuPc molecule.

In contrast to the band calculation stands the molecular energy level scheme, from which we know that the first electron addition and removal states are {\it not} the Cu $b_{1g}$ states, but the Pc derived $e_g$ and $a_{1u}$ states, respectively. It is very unlikely in the solid this level scheme is altered very much. Rather, the different results in the solid are due to the fact that LSDA treats electron correlations between localized electrons insufficiently. Here we are dealing with molecules that have an open $d$ shell and electron-electron correlation effects are known to be very strong for the Cu $d^9$ electrons. To include these effects one needs to go beyond LSDA. 

{\it Pristine CuPc: LSDA+U} We incorporate the electron correlations that are due to the copper ions by an effective Coulomb interaction $U_{Cu}$ in the $LSDA+U$ calculation scheme~\cite{LDAU}. The parameter $U_{Cu}$ is determined independently by computing the variation with respect to occupation number of the energy eigenvalue of the $b_{1g}$ state for a single relaxed CuPc molecule, using the GAMESS code \cite{GAMESS}. We find using a TZV basis set and the B3LYP functional that $U_{Cu}=6.28$ eV, a value that is typical for Cu $d^9$. The presence of such strong interaction effects is in accordance with the recent calculations of Tosatti {\it et al.}~\cite{Tosatti04}, which emphasize the importance of strong electron correlations. Note that the parameter $U_{Cu}$ that we calculated includes screening of the Coulomb interaction by the electrons on the Pc-ring of the molecule. In a crystal there will be additional screening of $U_{Cu}$ by the surrounding molecules. This will effectively reduce $U_{Cu}$ \cite{Meinders95,Brink95,Brocks04}, but we find that our conclusions on the symmetry, character and width of the valence and conduction do not depend on such additional screening.

The LSDA+U bandstructure is shown in Fig.~\ref{fig:LSDA+U}. The system is magnetic, with an energy gain with respect to the paramagnetic state of 0.622 eV per molecule. Due to $U_{Cu}$ the flat $b_{1g}$-derived Cu band below the Fermi-level  is shifted downward in energy with respect to the Pc derived states and the empty $b_{1g}$ is shifted upward. Now both the HOMO and LUMO derived band is of Pc character, in accordance with the energy level scheme of the bare molecule.  We find a valence band width of 0.32 eV. The correct order of the bands is very important, because next we will consider the electron doped system, for which the character of the valence band (very flat Cu-derived or Pc-derived and dispersing) is essential.
From the calculations we find a bandgap of $\Delta_{band}=1.2$ eV. This is should be contrasted with molecular CuPc, which has an energy gap of $\Delta_{CuPc}= I_{CuPc}-A_{CuPc}=4.5 \pm 0.3$ eV~\cite{Rosa94,Liao01}, where $I$ and $A$ are the ionization potential and electron affinity, respectively. These two numbers, however, should only be compared when two additional effects are taken into account. First, due to the polarization of the surrounding molecules the gap in the solid is screened with respect to its molecular value~\cite{Meinders95,Brink95,Brocks04}. We calculated this screening energy to be $1.7 \pm 0.1$ eV, so that on the basis of this result one would expect to find a bandgap in the solid of about $\Delta \approx 2.8 \pm 0.3$ eV. Thus, LSDA+U underestimates the gap. It is obvious that this is due to the fact that in such a calculation only the electronic correlations for the Cu-electrons are taken into account. The electron-electron interactions of the electrons on the Pc-ring, which makes out the largest part of the molecule, are not properly treated. We therefore determined with GAMESS the Coulomb interaction, the Hubbard $U_{Pc}$, for electrons on the Pc-ring of the molecule. We find a value of $U_{Pc}=3.0$ eV. Taking this into account we find a charge gap on the basis of the band calculations of $\Delta \approx 1.2 +3.0 -1.7=2.5 \pm 0.1$ eV. Thus within the error bars the two estimates of the gap agree and they are also consistent with the reported experimental values from photoemission spectroscopy ($2.3 \pm 0.4$ eV~\cite{Hill00}), transport properties ($2.1\pm 0.1$ eV~\cite{Schwieger02}) and the lower bound on the gap of 2.1 eV~\cite{Schwieger02} that is determined by energy electron loss spectroscopy.

\begin{figure}
\includegraphics[width=1.0\columnwidth]{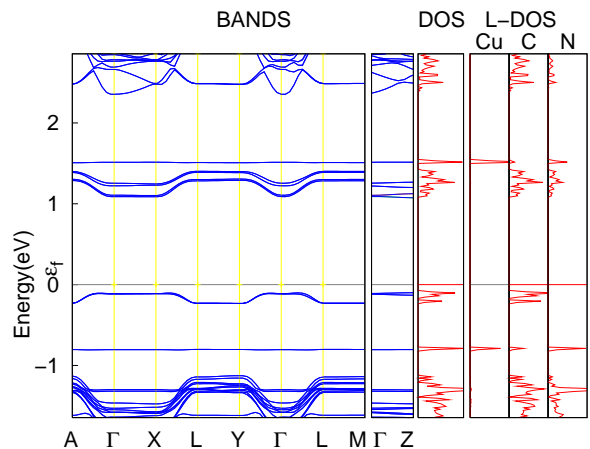}
\caption{
Band-structure, density of states and projected density of states for CuPc, calculated with SGGA+U.
}
\label{fig:CuPc_LSDA+U}
\end{figure}

{\it Doped Cu-phthalocyanine} We used the structural refinement of potassium intercalated CuPc reveals that the $K_{2.75}CuPc$ forms a stable phase. Its unit cell is monoclinic, has $C_{2h}$ space-group symmetry, and contains two CuPc molecules~\cite{Craciun03}. 
The structure is quite different from pristine CuPc, because now there is one molecule in the unit cell at $(0,0,0)$, as before, but the other one in the center of the cell, at $(a/2,b/2,c/2)$. According to the X-ray data there are two inequivalent positions of the potassium ions, which are both partially occupied. To perform calculations, we initially occupy only one of the two equivalent K-positions -the so-called $\rm K_{29}$- so that we have four potassium ions per unit cell, i.e. we start by analyzing ${{K}_2}{CuPc}$ before looking at the electronic properties of $\rm K_{2.75}CuPc$. There are two reasons to occupy only one K-position. First, when investigating the electronic structure of our material we want to separate its intrinsic electronic properties from the ones induced by disorder on the K-sites. Second, the present unit cell is already very large (118 atoms) and the additional inclusion of disorder (for instance in a super-cell or coherent potential approximation method) is numerically too demanding.  We will see the band-structure of $\rm{{K}_2}{CuPc}$ forms the proper starting point to understand the electronic structure of the higher doped $\rm{{K}_{2.75}}{CuPc}$ compound.

\begin{figure}
\includegraphics[width=0.8\columnwidth]{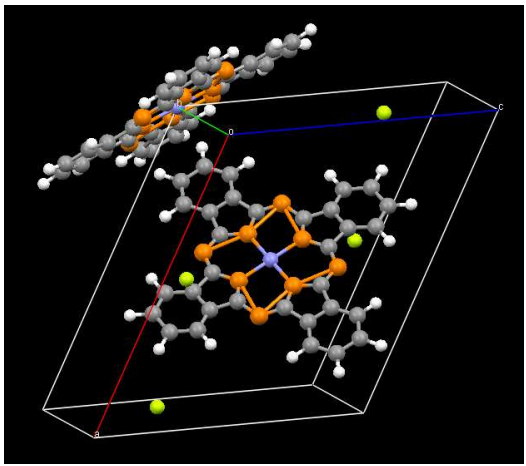}
\caption{Schematic representation of the crystal structure of $\rm K_{2.75}CuPc$.
}
\label{fig:structure}
\end{figure}

\begin{figure}
\includegraphics[width=1.0\columnwidth]{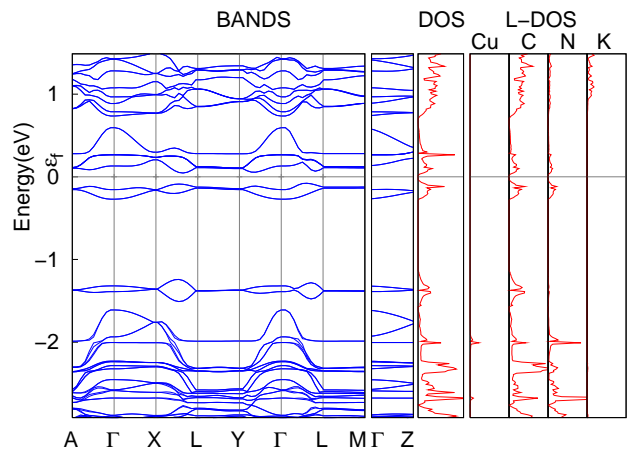}
\caption{
Band-structure, density of states and projected density of states for $\rm {(K_{2}{CuPc})}_2$, calculated with SGGA+U. The band energies are plotted along lines in the first Brillouin zone, connecting the points in the Brillouin zone as before.}
\label{K2CuPcSGGAU}
\end{figure}

Before being able to calculate the full bandstructure of $\rm{{K}_2}{CuPc}$ , we need to determine its groundstate properties. To this end, we first optimize the total energy allowing the ionic positions to relax. We observe that the K-ions move slightly in the unit cell to a position that is directly above the center of a phenyl ring of the phthalocyanine molecules --a high symmetry position. In addition, we have also looked at the magnetic structure of the ground state.  Again, these calculations are done with a kinetic cut-off energy of $550$ eV and the tetrahedron with Bloch correction, using 40 irreducible k-points. The results of the $SGGA+U$ band calculation on antiferromagnetic $\rm{{K}_2}{CuPc}$ and the DOS are shown in Fig.~\ref{K2CuPcSGGAU}. 

We see that the DOS of the valence and conduction bands does not have any K-character. We thus conclude that each potassium atom fully donates an electron to the CuPc rings so that we are dealing with $\rm CuPc^{2-}$. Apart from this, the K's do not play any role in determining the electronic properties. This implies that for a qualitative understanding of the electronic structure of potassium doped CuPc, the electronic structure of a CuPc molecule is a reasonable starting point.

We focus in Fig.~\ref{K2CuPcSGGAU}on the five bands around the Fermi level: the two that make up the valence band below it and the three of the conduction band above. 
Compared to pristine CuPc the valence band has taken up two additional electrons per molecule. This valence band is entirely derived from molecular $e_g$ orbitals, states that are localized on the Pc-ring. The  $e_g$ orbitals of a neutral CuPc molecule can absorb up to four electrons because these states are two-fold degenerate (besides spin degenerate).  So for $\rm{{K}_2}CuPc$ each CuPc molecule is left with two holes in the $e_g$ orbital. These empty states form two out of the three conduction bands. The third conduction band is the derived from the empty molecular $b_{1g}$ states, which is centered on the copper ions and has very little dispersion. 

From Fig. \ref{K2CuPcSGGAU} we see that the valence and conduction band, which are derived from the molecular $e_g$ states, are separated by a small gap of 110 meV. The dispersion of valence and conduction band is due to the hopping of electrons between neighboring molecules.  We find that the Pc-derived $e_g$ conduction band and the empty $b_{1g}$ Cu-band are crossing but that there is very little hybridization. 

Apparently the $e_g$ bands are also split, giving rise to a band gap. This is related to the fact that, as we already noted above, in a free CuPc molecule the states $e_g$ symmetry are two-fold degenerate \cite{Liao01}. In the $\rm {{K}_2}CuPc$ crystal this degeneracy is lifted due to the ionic and covalent molecular crystal fields. The 110 meV band splitting between the Pc-derived valence and conduction band, however, is only a fraction of the factual charge gap of $\rm {{K}_2}CuPc$. As in the pristine system we have to take the Coulomb interactions between electrons on the Pc-ring ($U_{Pc}$ and its screening) into account in order to find the correct gap value. After doing so we find $\Delta=1.4 \pm 0.1$ eV.  

\begin{figure}
\includegraphics[width=1.0\columnwidth]{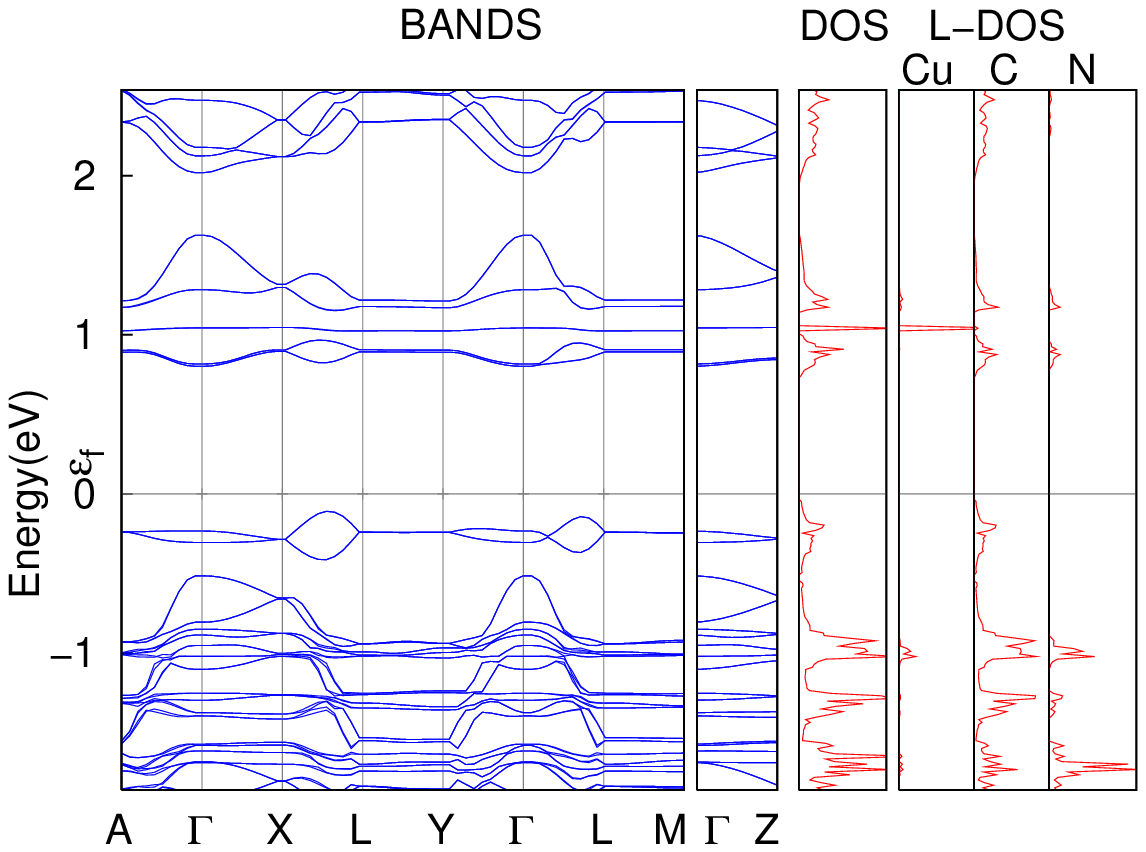}
\caption{
Band-structure, density of states and projected density of states for $\rm {(K_{0}{CuPc})}_2$, calculated with SGGA+U.
}
\label{KoCuPcSGGAU}
\end{figure}

In order to check how robust these results are, we have performed the same calculations and analysis for an identical crystal structure, but with all the potassium ions removed. We refer to this as the $\rm {{K}_0}CuPc$ structure. In this case we find a very similar band-structure to the one of $\rm {{K}_2}CuPc$, only with the Fermi-level shifted (Fig. \ref{KoCuPcSGGAU}). Apart from this shift, the only appreciable change in band-structure is that doping induces a small shift of $150$ meV to higher energy of the almost dispersionless Cu $b_{1g}$ band. From this we conclude that potassium intercalation gives rise to a simple band filling of the CuPc-related electronic states.  

The rigid band behavior is important when we consider further doping of electrons in $\rm {K_{2.75}{CuPc}}$ as in this compound we have for each CuPc molecule an extra of $0.75$ electrons, compared to $\rm {K_{2}{CuPc}}$. We calculated that the extra $0.75$ electrons shift up the Fermi-level in Fig. \ref{K2CuPcSGGAU} by $0.154$ eV. This implies that the doped carriers start filling the $e_g$ Pc-derived conduction band and the Cu-derived $b_{1g}$ band stays empty: the occupation of the $b_{1g}$ orbital turns out to be  $0.0152$ electron per CuPc. From this we conclude that electronic transport properties of $\rm {(K_{2.75}{CuPc})}$ are fully determined by the Pc-derived $e_g$ states. 

{\it Conclusions}
We find that pristine CuPc is a magnetic insulator with strong electron correlations, a charge gap of $2.5 \pm 0.1$ eV and a Pc-derived valence band with a width of 0.32 eV. $\rm {(K_{2}{CuPc})}$ is a magnetic insulator as well, with a gap of $1.4 \pm 0.1$ eV that is almost completely due to molecular Coulomb interactions. Our most important result concerns the bandwidth of the conduction band of $\rm {(K_{2}{CuPc})}$. It is $0.56$ eV, which is large for a molecular solid.  As, due to the rigid band behavior, in $\rm {(K_{2.75}{CuPc})}$ the additional $0.75$ electron per CuPc molecule fill the $e_g$ Pc-derived conduction band, this wide and partially filled band entirely determines electronic transport. On the basis of these band structure calculations one expects that $\rm {(K_{2.75}{CuPc})}$ is a metal. Note that in molecular crystal polaronic effects and disorder, not included in our analysis, tend to localize charge carriers, especially in the low doping regime. However in $\rm {(K_{2.75}{CuPc})}$ the large amount of doped charge carriers enables an effective screening of disorder, and the unusually large bandwidth makes polaron formation less likely, so that delocalization of the carriers is strongly promoted. 

{\it Acknowledgments} We thank Monica Craciun, Serena Margadonna and Alberto Morpurgo for stimulating discussions and for sharing experimental data with us prior to publication.

\end{document}